\documentclass[12pt,reqno,oneside]{amsart}

\usepackage{amsfonts,amssymb,latexsym}  
\usepackage{amsmath}  
\usepackage{natbib}
\usepackage{graphicx}
\usepackage{graphpap}
\usepackage{afterpage}

\newcommand{\N}{\mathbb{N}}

\renewcommand{\P}{\mathbb{P}}    
\newcommand{\PA}{\mathcal{P}_A}

\newcommand{\HH}{\mathcal{H}}

\newcommand{\given}{\ \vert \ }

\numberwithin{equation}{section}

\theoremstyle{definition}
\newtheorem{definition}[equation]{Definition}

\begin{document}

%%%%%%%%%%%%%%%%%%%%%%%%%%%%%%%%%%%%%%%%%%%%%%%%%%%%

\title{Detecting phylogenetic relations out from sparse context trees}
 
 \author{Florencia Leonardi}
 \address{Instituto de Matem\'atica e Estat\'istica, Universidade
  de S\~ao Paulo.\\ Rua do Mat\~ao 1010 CEP 05508-090, S\~ao Paulo, SP, Brazil.
   }
\email{leonardi@ime.usp.br}

\author{Sergio R. Matioli}
\address{Instituto de Bioci\^encias, Universidade de S\~ao Paulo.\\
   Rua do Mat\~ao, trav. 14, nº 321 CEP 05508-900,  S\~ao Paulo, SP, Brazil.\\
 }
 \email{srmatiol@ib.usp.br}

\author{Hugo A. Armelin}
\address{Instituto de Qu\'imica, Universidade de S\~ao Paulo.\\
   Av. Prof. Lineu Prestes, 748 CEP 05508-900,  S\~ao Paulo, SP, Brazil.
 }
 \email{haarmeli@iq.usp.br}

\author{Antonio Galves}
\address{Instituto de Matem\'atica e Estat\'istica, Universidade
  de S\~ao Paulo.\\ Rua do Mat\~ao 1010 CEP 05508-090, S\~ao Paulo, SP, Brazil.\\
   }
\email{galves@ime.usp.br}

\begin{abstract}
 The goal of this paper is to study the similarity between sequences
  using a distance between the \emph{context} trees associated to the
  sequences. These trees are defined in the framework of \emph{Sparse
    Probabilistic Suffix Trees} (SPST), and can be estimated using the
  SPST algorithm.
 We implement the Phyl-SPST package to compute the distance between 
 the sparse context trees estimated with the SPST algorithm.  
 The distance takes into account the structure of the  trees, 
 and indirectly the  transition probabilities. We apply this
  approach to reconstruct a phylogenetic tree of protein sequences in the globin
  family of vertebrates. We compare this tree with the one obtained using the well-known PAM distance.
  \end{abstract}

\maketitle

\section{Introduction}

In this work we propose to use the framework of Sparse Probabilistic Suffix Trees
(SPST) to analyze the similarity between sequences and to infer the
evolution of protein families. SPST was first introduced in
\citet{leonardi2005a} as a generalization of the PST algorithm,
proposed in \citet{ron1996}. SPST has shown to be useful in protein
modeling and classification, performing better than the PST algorithm
\citep{leonardi2006a}. The model that inspired the SPST algorithm is 
a generalization of Variable Length
Markov Chains (VLMC), introduced by \citet{rissanen1983}, and takes
into account the property of sparseness of the sequences. Given a
sequence, SPST estimates a set of \emph{sparse contexts}. A sparse
context is a short sequence of sub-sets of symbols (in a given alfabet) that are
relevant to predict any symbol in the sequence, given that the preceding
symbols belong to the sub-sets of the context. The SPST algorithm also
estimates the transition probabilities associated to each context. The
transition probabilities give the probability of each symbol
conditioned on the fact that the preceding symbols belong to the
sparse context. 

An interesting property of the set of sparse contexts is that it induces a partition of the
set of all possible sequences and can be represented as a tree. We use this
partition property to define a distance between context trees. This
distance can be used to measure the similarity between protein
sequences.

To our knowledge it has not been
proposed yet in the literature a method for sequence comparison using
the information contained in the architecture of the context trees associated to the
sequences. The more closely related approaches proposed until date are those that model the 
sequences as first order Markov chains and use a statistical measure to infer the similarity between
 them \citep{wu2001,pham2004}. The more remarkable difference between these approaches and our is 
that we do not use directly the estimated probabilities of the model. Instead of that we use the context 
tree architecture, that is trivial in first order Markov chains.  We show here that the context tree 
 architecture can have important structural information that may be useful to measure the similarity 
between sequences.

The paper is organized as follows. In Section~2 we review some
definitions in the framework of SPST. In Section~3 we introduce the
distance between sparse trees. In Section~4 we present the results
obtained for the globin protein family of vertebrates and finally in Section~5 we
discuss some aspects of our method.

\section{Sparse Context Trees}

Let $A$ be a finite alphabet (for example, the set of twenty amino
acids) of size $|A|$. We will denote by 
$\PA$ the set of parts of $A$. That is,
\[
\PA = \{v\colon v\subset A\}.
\] 
The elements in $\PA^j$ will be denoted by 
$w=(w_{-j},\dotsc,w_{-1})$. On the other hand, we will denote by $\PA^*$ 
the set of all finite sequences of elements in $\PA$; that is,
\[
\PA^* = \bigcup_{j=1}^\infty \PA^j.
\]

\begin{definition}\label{def:parsi}
  Let $(X_t)_{t\in\N}$ be a stochastic process taking values on the 
  finite alphabet $A$. We will say that the process $(X_t)_{t\in\N}$ is a 
  \emph{sparse stochastic chain} if there exists a set $\tau
  \subset \PA^*$ such that:
  \begin{enumerate}
  \item For any sequence $x_0,\dots,x_n$ satisfying 
    \[
    \P[X_0=x_0,\dotsc, X_{n-1}=x_{n-1}] > 0, 
    \]
    there exists an element $(w_{-k},\dotsc,w_{-1})\in\tau$ such that 
    \begin{align}\label{eq:prob}
      \P[X_n&=x_n | X_{n-1}=x_{n-1},\dotsc,X_0 = x_0] =\notag\\[1pt]
      &\P[X_n=x_n | X_{n-1}\in w_{-1},\dotsc,X_{n-k}\in w_{-k}].
    \end{align}
  \item If $(w_{-k},\dotsc,w_{-1})$ and $(\bar w_{-\bar k},\dotsc,\bar
    w_{-1})$ belong to $\tau$ and there exists $j$ such that $w_{-i}\cap
    \bar w_{-i}\ne \emptyset$ for $i=1,\dotsc,j$, then $w_{-i} =
    \bar w_{-i}$ for $i=1,\dotsc,j$.
  \item The set $\tau$ is the \emph{minimum} that satisfies 1. and 2. That is;
   if  $\bar\tau$ satisfies 1. and 2. then, for any $(\bar w_{-\bar
      k},\dotsc,\bar w_{-1})\in\bar\tau$ there exists 
    $(w_{-k},\dotsc,w_{-1})\in\tau$ such that $\bar k\geq k$ and $\bar
    w_j\subset w_j$ for all $j=1,\dotsc,k$.
  \end{enumerate}
\end{definition}

Each sequence $(w_{-k},\dotsc,w_{-1})\in\tau$ is called \emph{sparse context}
and the set $\tau$ is called \emph{sparse context tree}. This name is justified 
because  the set of sparse contexts can be represented as a rooted tree. 
In this tree, each context $w = (w_{-k},\dotsc,w_{-1})$ is represented by a complete
branch, in which the first node on top is $w_{-1}$ and so on until the
last element $w_{-k}$ which is represented by the terminal node of
the branch (Fig.~\ref{trees}). 

\setlength{\unitlength}{1mm}

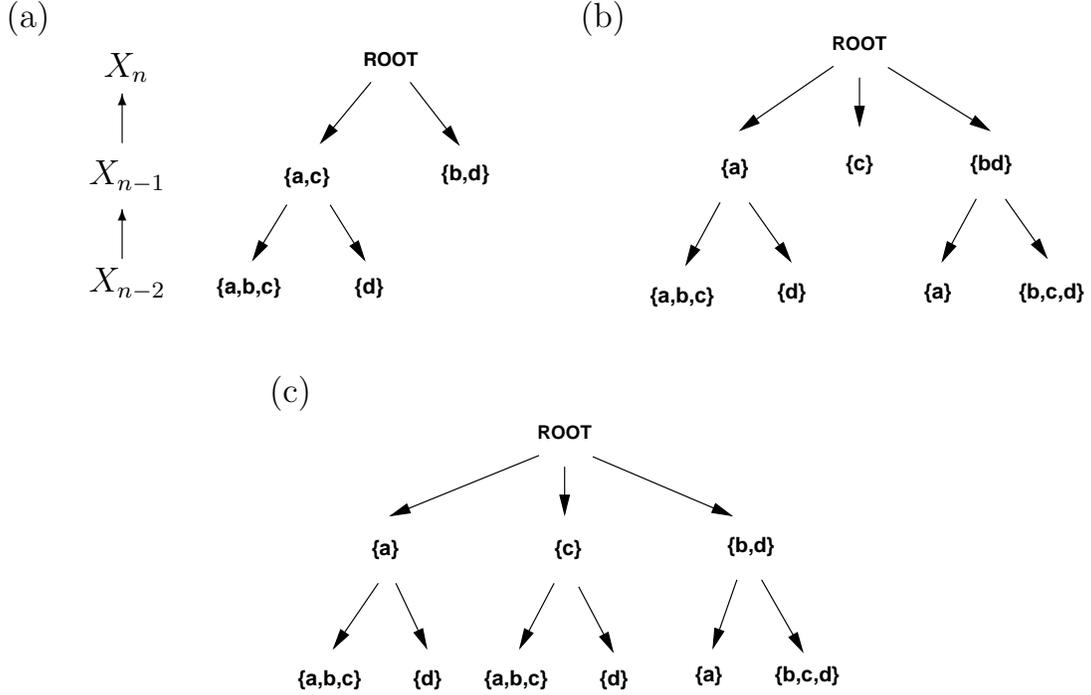
\begin{figure}[t]
\begin{center}
\scalebox{1.1}{
       \begin{picture}(70,45)(0,-5)
   %\graphpaper[2](0,-5)(70,45)
    \put(0,38){(a)}
    \put(24,5){
      \scalebox{0.43}{\includegraphics{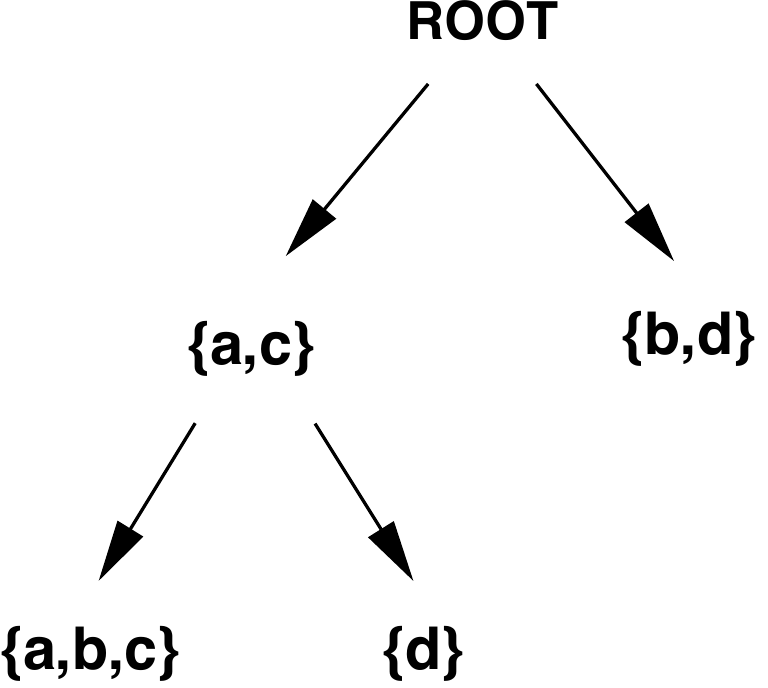}}
      }
      \put(10,6){$X_{n-2}$}
      \put(14,10){\vector(0,1){6}}
      \put(10,19){$X_{n-1}$}
      \put(14,24){\vector(0,1){6}}
      \put(11.8,32){$X_n$}
      \end{picture}
       \begin{picture}(70,45)(0,-5)
    %\graphpaper[2](0,0)(70,40)
      \put(-2,38){(b)}
     \put (5,4){
     \scalebox{0.43}{\includegraphics{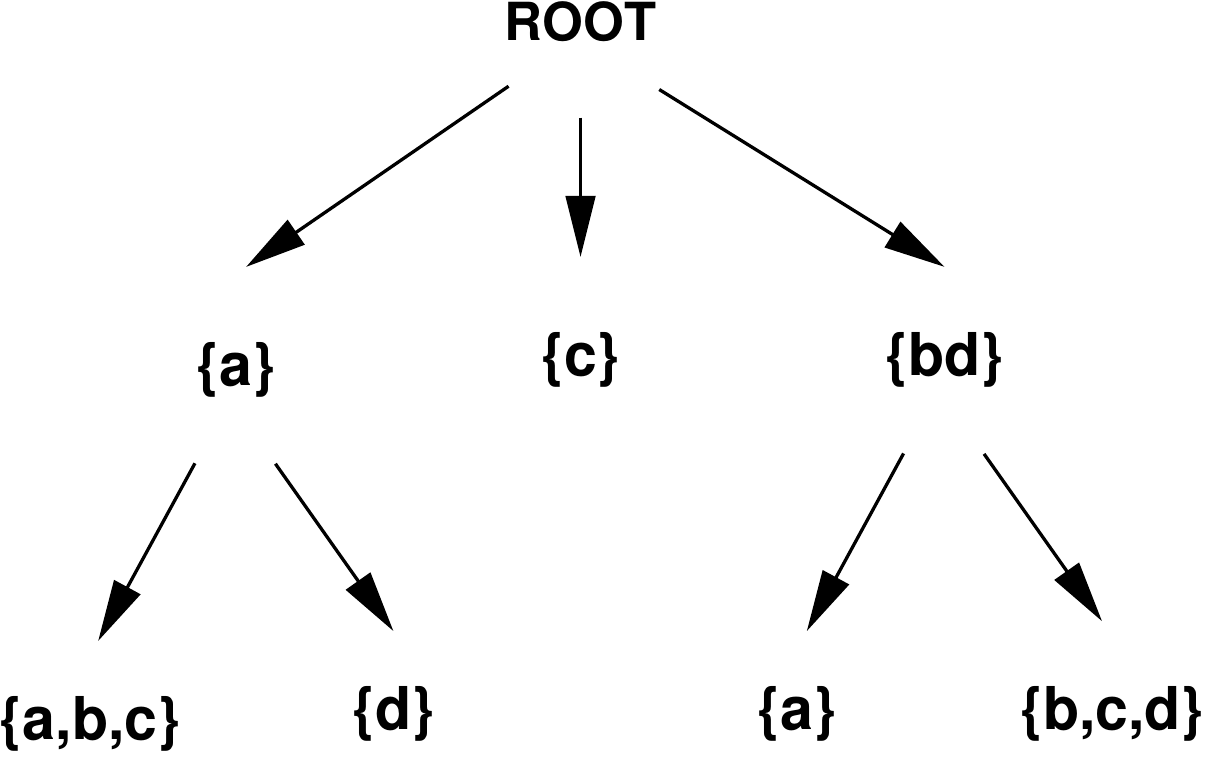}}
    }
     \end{picture}
     }
     \scalebox{1.1}{
      \begin{picture}(70,37)(0,3)
   % \graphpaper[2](0,3)(70,37)
      \put(0,38){(c)}
     \put(2,3){
       \scalebox{0.5}{\includegraphics{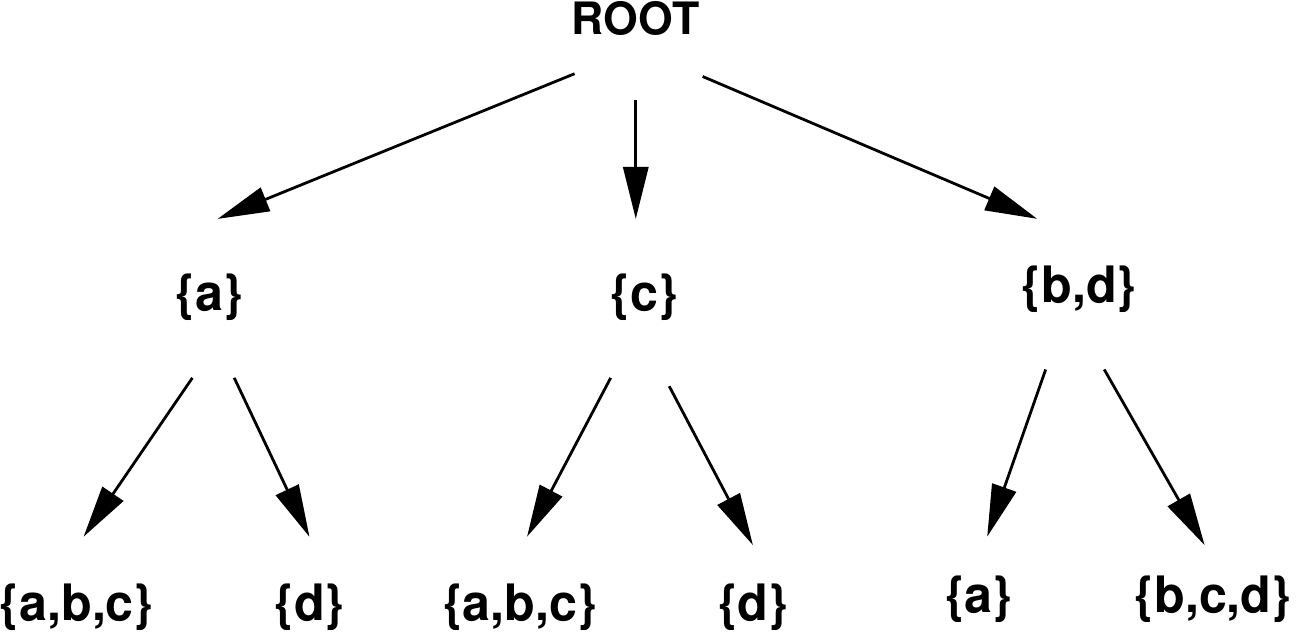}}
     }  
  \end{picture} 
  }
\end{center}
  \caption{Examples of sparse trees over the alphabet $A = \{a,b,c,d\}$. (a) The index of the variables 
   grows in the direction from the leaves to the root. In this case, the set of sparse contexts is
    $\{(\{a,b,c\},\{a,c\}),(\{d\},\{a,c\}),(\{b,d\})\}$.  (c) Maximum between the trees in (a) and (b).
   }
  \label{trees}
\end{figure}

Recently, it was proposed an algorithm to estimate the set of sparse
contexts and the transition probabilities given by \ref{eq:prob} \citep{leonardi2005a,leonardi2006a}. This
algorithm represents internally the set of sparse contexts as a tree,
as described above. 
We believe that this tree
contains important structural information that can be used to measure
the similarity between sequences. Our goal in this paper is to show 
some results concerning  this
conjecture. With this aim we propose to use a distance between
sparse context trees to measure the relatedness between
symbolic sequences. This distance is defined in the next section.

\section{A metric space of sparse trees}

Given a sparse context $w = (w_{-k},\dotsc,w_{-1})$ we denote
by $l(w)$ its length, that is $l(w) = k$. We use the
notation $s(w)$ for the product of the cardinals of the $w_i$'s, that is
\[
 s(w) \,=\, \prod_{i=1}^{l(w)} |w_i|,
\]
where $|w_i|$ is the number of symbols in $w_i$.

Given two sparse contexts $w = (w_{-k},\dotsc,w_{-1})$ and $\bar w =
(\bar w_{-\bar k},\dotsc,\bar w_{-1})$ we define the intersection between $w$
and $\bar w$ (assuming without loss of generality that $k\geq\bar k$) by
$w\cap \bar w = (w_{-k},\dotsc,w_{-(\bar k+1)},w_{-\bar k}\cap
\bar w_{-\bar k},\dotsc,w_{-1}\cap \bar w_{-1})$, if $w_i\cap\bar w_i \neq
\emptyset$ for all $i=1,\dotsc,\bar k$. In the case $w_i\cap\bar w_i =
\emptyset$ for some $i=1,\dotsc,\bar k$ we define $w\cap \bar w =
\emptyset$.

Given two sparse trees $\tau = \{w^1,\dotsc,w^n\}$ and $\bar\tau =
\{\bar w^1,\dotsc,\bar w^m\}$, we define the maximum between $\tau$ and $\bar\tau$
by
\begin{equation*}
  \tau\vee\bar\tau = \{w^i\cap \bar w^j \given w^i\cap \bar w^j \neq \emptyset;\, i=1,\dotsc,n;\, j=1,\dotsc,m\}.
\end{equation*}
The maximum between the trees of Figure~\ref{trees}(a)-(b) can be seen in Figure~\ref{trees}(c).

Before defining the distance between sparse context trees we introduce the 
notion of $\beta$-entropy of a tree $\tau$. Following
\citet{simovici2006} we define, for all $\beta>0$,
\begin{equation*}
  \HH_\beta(\tau) = \frac{1}{2^{1-\beta}-1}\,\Bigl(\;\sum_{w\in\tau}
  \bigl[s(w)\,|A|^{-l(w)}\bigr]^\beta - 1\,\Bigr),\quad\text{if $\beta\ne 1$,}
\end{equation*}
and 
\begin{equation*}
  \HH_\beta(\tau) = - \sum_{w\in\tau} s(w)\,|A|^{-l(w)} \cdot \log_{2} \bigl[ s(w)\,|A|^{-l(w)}\bigr],\quad\text{if 
$\beta = 1$.}
\end{equation*}
Then, given two sparse trees, $\tau$ and $\bar\tau$, we define the $\beta$-distance between them as 
\begin{equation}\label{eq:dist}
  d_\beta(\tau,\bar\tau) = 2\,\HH_\beta(\tau\wedge\bar\tau) - \HH_\beta(\tau) - \HH_\beta(\bar\tau).
\end{equation}
It can be seen that $ d_\beta(\cdot,\cdot)$ defines a distance over the set of all context trees. 
The proof of this assertion can be found in \citet{simovici2006}.

\section{Results}

We implemented an algorithm coded in C, called Phyl-SPST, to calculate
distances between context trees, as defined by (\ref{eq:dist}).  The
source code and compiled versions for Mac OS X, Linux/Unix and Windows
can be downloaded from the site
{\ttfamily http://www.ime.usp.br/numec/softwares/phyl-spst/}.

 \begin{figure}[t]
  \begin{center}
   \includegraphics[width=10cm,trim = 0 25 0 10]{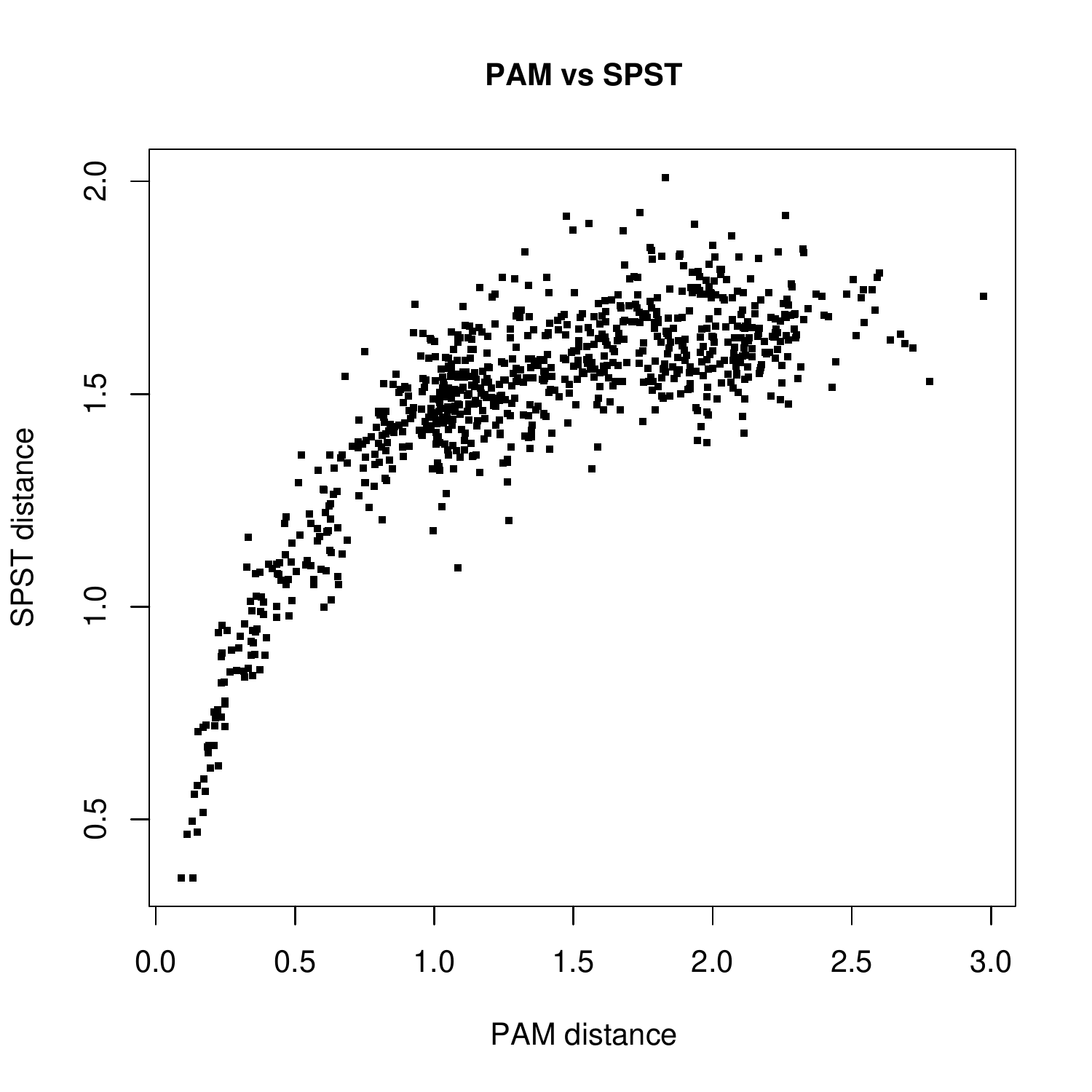}
  \end{center}
  \caption{Comparison of the SPST and PAM distance matrices.}
  \label{fig:distancias}
\end{figure}

\begin{figure}[h!]
  \begin{center}
   \flushleft{(a)}
     \begin{center}
    \scalebox{.81}{\includegraphics{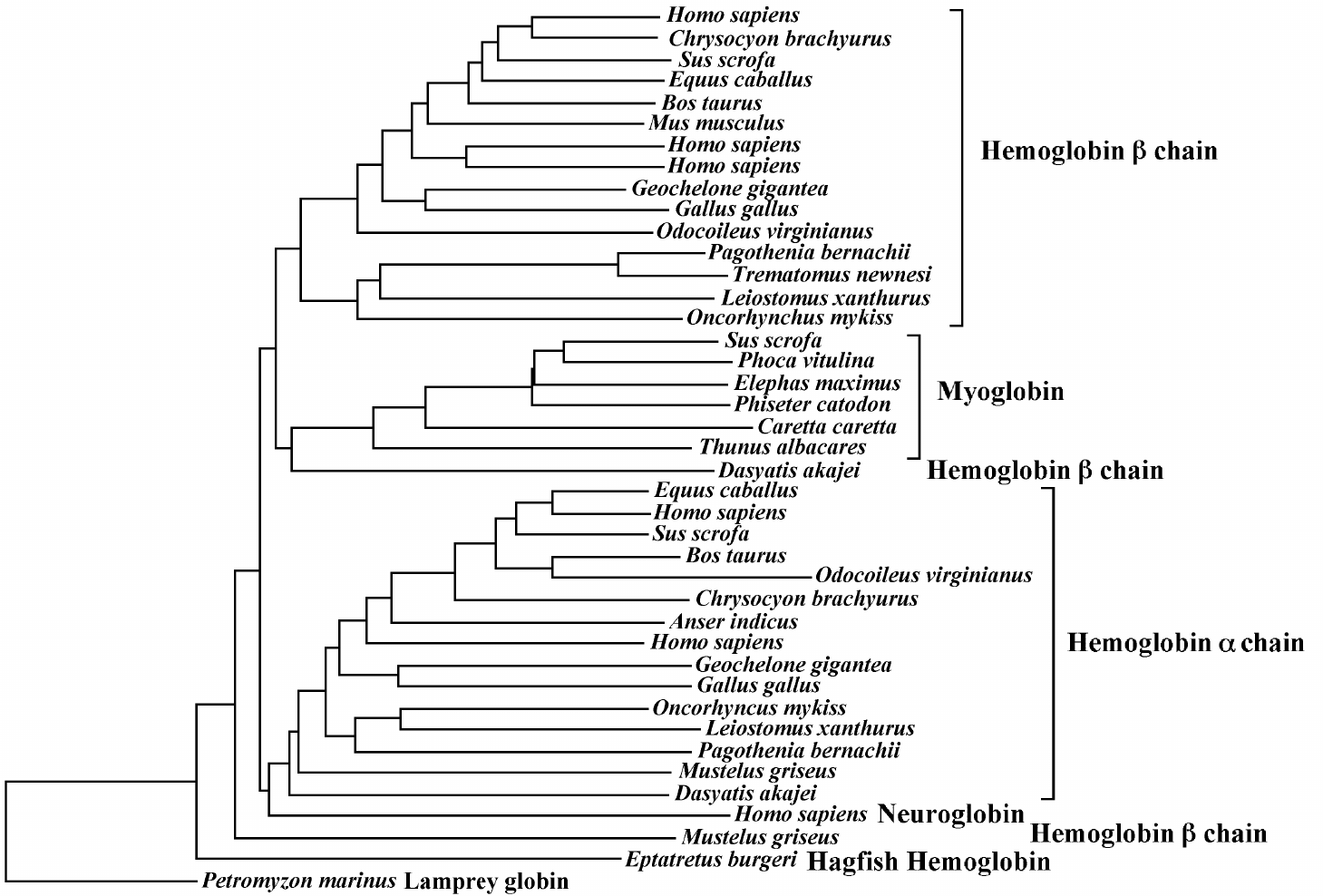}}
     \end{center}
     \vspace{.5cm}
  \flushleft{(b)}
  \begin{center}
  \scalebox{.81}{\includegraphics{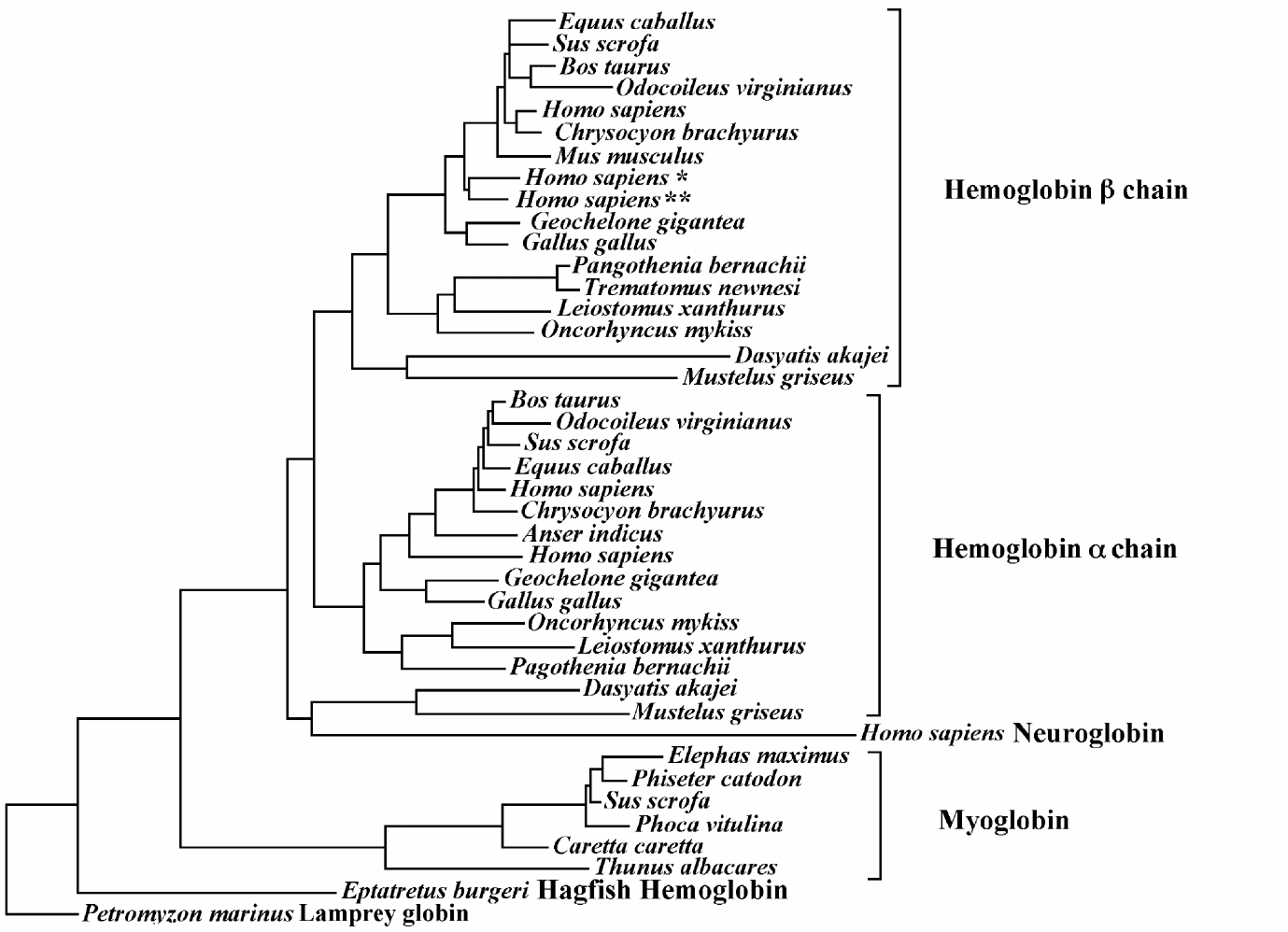}}
\end{center}
    \end{center}
    \caption{Phylogenetic trees made with Neighbor Joining clustering
      algorhitm on SPST distances (a) and on PAM distances (b) }
     \label{globinas}
  \end{figure}

  We applied the Phyl-SPST package to study the similarity between the
  protein sequences of the globin family of vertebrates.  The 41
  sequences used in this analysis were obtained from the SCOP database
  \citep{scop2004} and can be found in the supplementary material. The
  program estimated, for each sequence in this set, a sparse context
  tree. Then it computed the distance matrix using the
  $\beta$-distance defined by (\ref{eq:dist}). In what follows we
  call this distance the SPST distance. In order to compare our
  method with an alignment-based distance we used the structure based
  alignment of the 41 globin sequences of vertebrates present in the PALI
  database \citep{pali2003} (alignment available in
  supplementary material). Then, we applied the algorithm PROTDIST of
  the Phylip3.65 package \citep{phylip}, with the Dayhoff PAM matrix option,
  to compute the distance matrix.

  When the PAM and SPST distances are plotted against each other (Fig.
  \ref{fig:distancias}) a non linear relation is clearly observed. 
  With  each distance matrix we reconstructed a phylogenetic tree using the
  NEIGHBOR and DRAWGRAM algorithms of the Phylip3.65 package. These
  phylogenetic trees can be seen in Figure~\ref{globinas}.  In both
  trees the lamprey globin was used as outgroup.

\section{Discussion}
The dataset we used to verify the potential use of the SPST distances on phylogenetic reconstruction is 
a vertebrate subset of the globin gene family. This family is one of the first protein families that was 
characterized  \citep{dayhoff1972} and is, perhaps, the most known to date \citep{vinogradov2006}. 
Besides, the vertebrate phylogeny is also well studied and is ground in relatively abundant 
paleontological, morphological, molecular, and physiological analyses \citep{cotton2002}.  

The phylogenetic tree shown in Fig. \ref{globinas}(a) proves that in fact the context trees 
inferred from symbolic sequences (in this case, protein sequences) can offer important   
evolutionary information of the sequences.  This constitutes an original and very promise aspect of the modeling of sequences by variable memory stochastic processes, and it needs to be studied in more details.

The phylogenetic analysis here performed also reflects the overall behavior of the SPST distance. The 
tree produced with the SPST present larger branches in the most inclusive sequences, and shorter 
branches in the most basal sequences. With respect to the tree topology, the main differences between 
them is the placement of the myoglobin cluster, that is closer to the beta chain of hemoglobin in the 
SPST tree and, in the PAM tree, it is outside of the hemoglobin chain. Other remarkable difference is the 
placement of the red tail deer (\emph{Odocoileus virginianus}) outside the cluster that contains the mammals, a reptile 
(\emph{Geochelone gigantea}), and a bird (\emph{Gallus gallus}) in the beta chain cluster of the SPST 
tree.  Although there are minor misplacements in the tree based on PAM distances with respect to the 
vertebrate and globin traditional phylogenies, it is superior in reconstructing the phylogeny than with the 
use of SPST distances.

The relationship between the SPST distance and the classical PAM distance of the globin family of 
vertebrates shows a plateau behavior. The short PAM distances yields larger SPST distances, and the 
opposite occurs when distances are longer.  This may be caused by the bounded nature of the context trees and by the specific form of the distance we propose. 
Therefore, this analysis shows that small differences in 
sequences causes enough changes in the context trees to increase the SPST distance between them. It remains yet as an open problem the characterization of the changes produced in the context trees by stationary modifications of the sequences as mutations, insertions or deletions. We think that these 
characterizations could help to improve the results shown here. On the other hand, it is also important to define and test other distances over the set of trees to study their specific behaviors and compare them to the one proposed here.

\section*{Acknowledgments}

This work is part of
PRONEX/FAPESP's project \emph{Stochastic behavior, critical phenomena
  and rhythmic pattern identification in natural languages} (grant
number 03/09930-9) and CNPq's project \emph{Stochastic modeling of speech}
(grant number 475177/2004-5).
During the preparation of this paper F.G.L was supported by a CAPES grant and by a FAPESP 
fellowship (process 06/56980-0). The authors A.G., H.A.A., and S.R.M. would like to thank 
the fellowship grants received from CNPq. The authors would like to thank Dr. Eleonora Trajano for comments on the vertebrates phylogenies.

\bibliographystyle{dcu}
\bibliography{references.bib}

\end{document}